\documentstyle[12pt]{article}
\newcommand{\beq}{\begin{equation}}
\newcommand{\eeq}{\end{equation}}
\newcommand{\bea}{\begin{eqnarray}}
\newcommand{\eea}{\end{eqnarray}}
\newcommand{\tr}{{\,\hbox{\rm Tr}\,}}
\newcommand{\la}{\left\langle}
\newcommand{\ra}{\right\rangle}
\renewcommand{\d}{{\, d}}

\newcommand{\stre}{{\sigma_3}}
\newcommand{\suno}{{\sigma_1}}
\newcommand{\euno}{{e^{i\phi}}}
\newcommand{\edue}{{e^{2i\phi}}}

\newcommand{\im}{{\,\hbox{\rm Im}\,}}

\newcommand{\rf}[1]{{(\ref{#1})}}

\begin{document}
\begin{titlepage}
\begin{flushright}
IFUM - 463/FT\\
March 1994\\
\end{flushright}
\vspace{0.5cm}
\begin{center}
{\Large {\bf
1D-Disordered Conductor with Loops\\
Immersed in a Magnetic Field}} \\
\vspace{1.5cm}
{\bf Alberto S. Cattaneo}
\footnote{E-mail: cattaneo@vaxmi.mi.infn.it}  \\
\vspace{0.4cm}
{\em Dipartimento di Fisica, Universit\`a degli Studi di Milano \\
and INFN, sezione di Milano, 20133 Milano, Italy.} \\
\vspace{0.4cm}
{\bf Andrea Gamba}
\footnote{E-mail: gamba@polito.it} \\
\vspace{0.4cm}
{\em Dipartimento di Matematica, Politecnico di Torino, 10129 Torino,
Italy \\
and INFN, sezione di Pavia, 27100 Pavia, Italy} \\
\vspace{0.4cm}
{\bf Igor V. Kolokolov}
\footnote{
Permanent address: Budker Institute of
Nuclear Physics, 630090 Novosibirsk, Russia.\\
E-mail: kolokolov@vaxmi.mi.infn.it, kolokolov@inp.nsk.su} \\
\vspace{0.4cm}
{\em Sezione INFN di Milano, 20133 Milano, Italy}

\vspace{0.4cm}
{\bf Maurizio Martellini}
\footnote{E-mail: martellini@vaxmi.mi.infn.it} \\

\vspace{0.4cm}
{\em Dipartimento di Fisica, Universit\`a degli Studi di Milano,
20133 Milano, Italy\\
and Sezione INFN di Pavia, 27100 Pavia, Italy.}
\end{center}
\vspace{2cm}

\begin{abstract}
\noindent
We investigate the conductance of a 1-D disordered
conducting loop with two
contacts, immersed in a magnetic flux.
We show the appearance in this model
of the Al'tshuler-Aronov-Spivak
behaviour.
We also investigate the case of a chain of loops
distributed with finite density:
in this case we show that the interference effects due to the
presence of the loops can lead to the delocalization of the wave
function.
\end{abstract}
\end{titlepage}
\newpage
\baselineskip=18pt
\section{Introduction}

The Aharonov-Bohm effect arises combining a non-trivial topology with the
presence of the magnetic
field~\cite{aharonov_bohm}.
When disorder is added to this picture it has been
shown~\cite{altshuler_etal}
that usually observed quantities acquire new properties.
For instance, the conductivity of a thin cylindrical conductor
immersed in a magnetic flux
$\phi$
is a periodic function
having period
$\phi_0$,
the quantum of magnetic flux,
but in the presence of disorder one gets a period
\(\phi_0/2\)
(see Ref.~\cite{altshuler_etal}).
This effect has been observed in
experiments~\cite{shavsin}.
In
Ref.~\cite{altshuler_etal}
computations were carried on for the 2D-case in the framework of
weak localization
theory~\cite{khmelnitskij_etal}
and essential use was made of perturbation theory.

The papers
\cite{asbel_etal,Bu,BuBu}
have arisen interest toward 1D-models presenting the
Al'tshuler-Aronov-Spivak behaviour.
In these works the study of the conductivity of a ring
immersed in a magnetic flux and having two contacts on opposite sides
was carried on.
The case of the ordered ring was studied there thoroughly.
In~\cite{li_soukolins}
it was considered a disordered ring, but localization effects
were completely ignored.

In the present paper we consider
a loop
of one-dimensional disordered conductor
immersed in a magnetic flux
$\phi$
(measured in units of
$\phi_0$).
The geometry of the contact that we are considering differs from the
one analyzed in
Refs.~\cite{asbel_etal}-\cite{li_soukolins}
(see section~2 and Fig.~1).
This allows us to describe the contact in terms of glueing matrices
for solutions of the Schr\"odinger equation rather than in terms
of the $S$-matrices used in
Refs.~\cite{asbel_etal}-\cite{li_soukolins}.
This point may turn useful in generalizations of the model.

We compute the conductivity of the sample in the framework of the
Landauer-B\"uttiker
approach
\cite{landauer,Bu}
and show that it is a periodic function of
$2\phi$,
so that the frequency of oscillations is doubled.
This result is in agreement with the considerations put forth
in
Ref.~\cite{altshuler_etal}
(see
also~\cite{altshuler_aronov} for a review).

\section{Conductance of a loop immersed in a magnetic flux}

In order to compute the conductance of the sample depicted in
Fig.~1 we resort to Landauer's formula
\cite{landauer,Bu}:
\beq\label{cond_a}
G={e^2\over 2\pi\hbar}{\tau\over1-\tau},
\eeq
where $\tau$ is the transmission probability (calculated at the
Fermi level).
$\tau$ can be obtained from the
$2\times 2$
transfer matrix
$T$
which maps the space of solutions
$\psi(x)$
of the Schr\"odinger equation from one space point
to another:
\beq\label{due}
v(x_2)=T \,v(x_1),
\qquad
v(x)=
\left(
\begin{array}{c}
\psi'(x) + ik\psi(x) \\
\psi'(x) - ik\psi(x) \\
\end{array}
\right),
\qquad
\tau=1/|T_{22}|^2;
\eeq
here
$k$
denotes the one-dimensional momentum of the electron
and the Schr\"odinger equation has the form
\beq\label{cinque}
-\psi''+U(x)\psi=k^2\psi.
\eeq
Let us determine the transfer matrix $T$ for a generic loop.
The one-dimensional description is adequate only
asymptotically, far away from the cross-junction domain.
Inside this domain we should use the three-dimensional Schr\"odinger
equation.
However, if the domain is small enough
({\it i.e.}, of size
$\sim 1/k$),
we just need to get some linear relations among
the effective asymptotic one-dimensional wave functions.
Since the Schr\"odinger equation is a second order PDE,
the fixing of the boundary values of the wave function
determines the solution uniquely with all its derivatives.

\setlength{\unitlength}{2.7pt}
\begin{figure}[t]
\vspace{36pt}
\begin{picture}(18,18)(-10,60)
\thicklines
\put(75.00,80.00){\circle{14.00}}
\put(75.00,73.00){\circle*{2.40}}
\put(75.00,73.00){\line(-1,-1){9.00}}
\put(75.00,73.00){\line(1,-1){9.00}}
\put(75.00,80.00){\circle*{1.89}}
\thinlines
\put(62.00,64.00){\line(1,0){8.00}}
\put(80.00,64.00){\line(1,0){8.00}}
\put(62.00,63.00){\line(1,0){8.00}}
\put(80.00,63.00){\line(1,0){8.00}}
\put(78.00,81.33){\makebox(0,0)[cc]{$\phi$}}
\end{picture}
\caption{\small The form of the conducting loop.}
\end{figure}
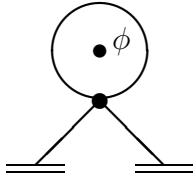

\setlength{\unitlength}{1.5pt}
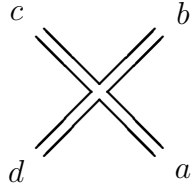
\begin{figure}[h]
\vspace{40pt}
\begin{picture}(26,26)(-80,60)
\thicklines
\put(75.00,80.00){\line(-1,1){14.00}}
\put(75.00,80.00){\line(-1,-1){14.00}}
\put(79.00,80.00){\line(1,1){14.00}}
\put(79.00,80.00){\line(1,-1){14.00}}
\put(77.00,82.00){\line(1,1){14.00}}
\put(77.00,82.00){\line(-1,1){14.00}}
\put(77.00,78.00){\line(1,-1){14.00}}
\put(77.00,78.00){\line(-1,-1){14.00}}
\put(98.00,100.00){\makebox(0,0)[cc]{$b$}}
\put(56.00,100.00){\makebox(0,0)[cc]{$c$}}
\put(56.00,60.00){\makebox(0,0)[cc]{$d$}}
\put(98.00,60.00){\makebox(0,0)[cc]{$a$}}
\end{picture}
\caption{\small The cross-junction.}
\end{figure}

\noindent
This means that among the eight quantities
$\psi(a)$, $\psi'(a)$,
$\psi(b)$, $\psi'(b)$,
$\psi(c)$, $\psi'(c)$,
$\psi(d)$, $\psi'(d)$
(see Fig.~2)
there should exist four linear relations, which can be written
for a symmetric junction as
\beq\label{junction}
\left(
\begin{array}{c}
v_a \\
v_b \\
\end{array}
\right)
=
\left(
\begin{array}{cr}
A & B \\
B & A \\
\end{array}
\right)
\left(
\begin{array}{c}
v_c \\
v_d \\
\end{array}
\right),
\eeq
with
$A$
and
$B$
being
$2\times 2$ complex matrices.
If we now connect $b$ and $c$ through a generic transfer matrix
$T_0$
(or
$e^{i\phi} T_0$
in the presence of a magnetic flux
$\phi$)
we get
$v_a$
in terms of
$v_d$
as follows:
\beq\label{transf}
v_a =T\,v_d,
\qquad
T    = B + e^{i \phi} A T_0 (1- e^{i \phi} B T_0)^{-1} A.
\eeq
Since $T$ is a transfer matrix, it must satisfy the current conservation
condition:
\beq\label{unitarity}
T^* T = 1,
\quad \mbox{\rm with} \quad T^*= \stre T^\dagger \stre,
\eeq
and the time-inversion invariance:
\beq\label{reality}
\tilde T = T \quad \hbox{\rm if} \quad \phi=0,
\quad \mbox{\rm with}\quad\tilde T= \suno \bar T \suno.
\eeq
The same equations hold for
$T_0$.
(\ref{unitarity})
and
(\ref{reality})
constrain the form of the matrices $A$ and $B$.
Eq.
(\ref{reality})
simply tells us that
$\tilde A = A$
and
$\tilde B = B$.
Notice that
$2\times 2$
matrices such that
$\tilde X = X$
have the form
$X = x_0 + \vec x\cdot\vec\sigma$,
where
$\vec\sigma$
is the vector of Pauli matrices,
$\vec x=(x_1,x_2, i x_3)$
and all the
$x_j$ are real.
For matrices of this kind the $*$-conjugation just amounts to a
``space'' reflection
$ X^* = x_0-\vec x\cdot\vec\sigma$.
Using these properties, and the fact that
(\ref{unitarity}) and (\ref{reality})
hold for
$T_0$,
we get the following inversion formula:
\beq\label{inversion}
(1-e^{i\phi} B T_0)^{-1} =
{1\over \Delta} (1-\euno T_0^* B^*),
\eeq
where
\beq\label{det}
\Delta=\det (1-\euno B T_0)=
1-\euno \tr (B T_0)+\edue \det B;
\eeq
thus
(\ref{transf})
can be rewritten as
\beq\label{transf_a}
T={1\over\Delta}
\left(
\Delta B + \euno  A T_0 A - \edue A B^* A
\right).
\eeq
We now plug
(\ref{transf_a})
into the unitarity condition~\rf{unitarity},
which must be satisfied by arbitrary values of
$\phi$
and
$T_0$,
and is thus equivalent to the vanishing of a matrix-valued
trigonometric polynomial.
Vanishing of the
$\sin 2\phi$-term
gives
\beq\label{main_a}
A^* B = - B^* A;
\eeq
for
$\phi=0$
and from the arbitrariness of
$T_0$
we get
\footnote{
Eq.
\ref{main_b}
can also be rewritten as
$B^* B + A^* A = 1$.
}
\beq\label{main_b}
\det A = 1 - \det B.
\eeq
The remaining terms give no additional conditions.

We remark that
(\ref{unitarity})
and
(\ref{reality})
are also satisfied
if we impose on
$A$
and
$B$
the equations which are obtained from
(\ref{main_a})
and
(\ref{main_b})
by reversing the signs of the right hand sides.
However, only the positive choice of the sign respect invariance
with regard to the interchange of $A$ and $B$.
This invariance has to be satisfied for the following reason:
if we connect $b$ and $d$ instead of $b$ and $c$ we would
obtain
(\ref{main_a})
and
(\ref{main_b})
with $A$ and $B$ interchanged.
It is easy to verify that the new equations would not be compatible
with the previous ones if we had chosen the negative signs.

Using
(\ref{main_a})
and
(\ref{main_b})
we can rewrite $T$ in the simpler form
\beq\label{transf_b}
T={\euno\over\Delta}
[(2\cos\phi-\tr B T_0)\cdot B + AT_0A].
\eeq
One can easily verify that the general solution of
(\ref{main_a})
and
(\ref{main_b})
depends upon six real parameters and is given by
\beq\label{solution}
B 	 = k S A
\eeq
with
\beq\label{sol_a}
S	 =
i \left(\begin{array}{cr}
	q	& z	\\
	-\bar z	& -q	\\
\end{array}\right),
\qquad \pm 1  = q^2 - |z|^2,
\qquad
k^2	 = \pm (1/\det A - 1),
\eeq
where
$A=a_0+\vec a\cdot\vec\sigma$
is an arbitrary matrix satisfying
$\tilde A = A$
and the positive signs in
(\ref{sol_a})
have to be chosen when
$0<\det A<1$.

Landauer's formula
(\ref{cond_a}), using
(\ref{transf_b}),
now reads
\beq\label{cond_c}
G={e^2\over 2\pi\hbar}{|\Delta|^2\over|Q|^2-|\Delta|^2},
\eeq
with
\beq\label{defq}
Q=(2\cos\phi-\tr B T_0)\cdot B_{22} + (AT_0A)_{22}.
\eeq

\section{Appearance of the AAS effect}

We shall consider the geometry depicted in Fig.~1, where the loop
of length $L$ is constituted by a 1-dimensional disordered
conductor.
In this case
the transfer matrix describing the propagation through the disordered
loop can be parametrized as
follows~\cite{dor}:
\beq\label{transf_c}
T_0 = \left(\begin{array}{cc}
 \alpha		& \beta		\\
 \beta^*	& \alpha^*
\end{array}\right)
 =
\left(\begin{array}{cr}
e^{i(\alpha_s+kL)} \cosh\Gamma	    & e^{i(\beta_s+kL)} \sinh\Gamma	\\
e^{-i(\beta_s+kL)}      \sinh\Gamma & e^{-i(\alpha_s+kL)}\cosh\Gamma\\
\end{array}\right),
\eeq
where
$\alpha_s$, $\beta_s$ and $\Gamma$
are functionals of the
impurities potential and can be considered as slow variables
(see~\cite{kol} for details).

The main point in the approach of Refs.
\cite{altshuler_etal,khmelnitskij_etal,altshuler_aronov,dor}
is the averaging over an ensemble of impurities potentials.
However, in the case of a given sample we also have to separate
fast-phase averaging and ensemble averaging.
Fast-phase averaging can be performed {\it e.g.} in the following way:
\beq
\label{reg}
G_L={1\over\Delta L}\int_L^{L+\Delta L} G\d L'
\qquad
(1/k\ll\Delta L\ll\ell),
\eeq
where $\ell$ is the localization length.
This average must be considered since the length
\( L \)
is defined with a precision
\(\Delta L\)
of the order of the size of the intermediate domain where the
wave function is essentially three-dimensional.
{}From~\rf{transf_c}
it follows that only functions of
\(|\alpha|^2\)
and
\(|\beta|^2\)
are non-vanishing after fast-phase averaging.

We are now in a position to prove the appearance of the
Al'tshuler-Aronov-Spivak
effect in our model.
{}From~\rf{det} and~\rf{defq}
it follows that the coefficients of
$\cos\phi$-terms
appearing in the r.h.s. of~\rf{cond_c}
are linear combinations of
$\alpha$, $\alpha^*$, $\beta$ and $\beta^*$.
Therefore after performing the average~\rf{reg} only even powers of
$\cos\phi$
survive and $G$ becomes a function of
period
$\pi$
instead of
$2\pi$.
An explicit example is given in the following section.
This shows that the conclusion
of Ref.~\cite{li_soukolins}
that at
$|\beta|\neq 0$
the
$\cos\phi$-oscillations
of the conductance survive after fast-phase averaging is wrong~\footnote{
We must stress here that the phases of the non-diagonal elements
of~\rf{transf_c} are fast variables.
}.

\section{Explicit computation}

Let us first of all consider the simple case
of the following 1-parameter solution of
(\ref{main_a})
and
(\ref{main_b}):
\beq\label{simplecase}
A = \sqrt{1-\kappa^2},	\quad B = i \kappa \stre.
\eeq
{}From~\rf{cond_c} we get
\beq\label{cond_d}
G={e^2\over 2\pi\hbar}
{4\kappa^2\im^2\alpha+4\kappa(1+\kappa^2)\cos\phi\im\alpha
+1+2\kappa^2\cos 2\phi+\kappa^4
\over
(1-\kappa^2)^2|\beta|^2}.
\eeq
After the fast-phase averaging we obtain
\beq\label{cond_e}
G_L	 = 	G_\infty+{e^2\over 2\pi\hbar}
		{1+4\kappa^2\cos^2\phi+\kappa^4
		\over
		(1-\kappa^2)^2}
		{1\over|\beta|^2},
\eeq
\beq
G_\infty =	{e^2\over 2\pi\hbar}
		{2\kappa^2\over (1-\kappa^2)^2}.
\eeq
In the general case the expressions of $G_L$ in terms of
$|\alpha|^2$, $|\beta|^2$ and $\cos^2\phi$ is rather cumbersome.
If $L/\ell\gg 1$
we have
\beq
|\beta|^2=\exp\,\log |\beta|^2 \sim\exp\la\log |\beta|^2\ra
\eeq
since the logarithm of
$|\alpha|^2$
and
$|\beta|^2$
in the limit
$L/\ell\gg 1$
is additive~\cite{lifshits_etal, melnikov}.
For large
$L/\ell$
the behaviour of
$\la\log|\beta|^2\ra$
is well
known:
\beq
\la\log|\beta|^2\ra\sim 2\,\frac{L}{\ell}.
\eeq
Thus,
the values of $|\alpha|^2$ and $|\beta|^2$
are large enough and $G_L$ can be represented as a series
expansion in $|\beta|^{-2}$.
At the first order we get
$G_L=G_\infty+C+D\,\cos^2\phi$
where $C$ and $D$ are of the order
$\exp(-2L/\ell)$.
This is in accordance with estimations~\cite{glasman}
made in the framework of weak-localization theory.

\section{Chain of loops}

In this section we want to show that the interference effects
due to the presence of loops can in principle lead to the delocalization
of the wave function.

The transfer matrix of a conducting loop
with coefficient of internal scattering
of order $\delta x$
({\it i.e.} with $A=1+A'\,\delta x$,
$B=B'\,\delta x$)
is obtained by linearizing~(\ref{transf_b}):
\beq
R=1+i\,\Theta\,\delta x
\eeq
where $\Theta$ is a matrix verifying
$\Theta^*=-\Theta$, $\tilde{\Theta}=\Theta$.
Let us consider a series of
$N$
such loops connected by short wires of disordered conductor of length
$\delta x$, and
let us denote by $x_i$ the position of the $i$-th loop.
The transfer matrix corresponding to the $i$-th loop is
$$
T_i=1+(i\,\varphi(x_i)\,
s_3+\zeta_+(x_i)\,s_-+\zeta_-(x_i)\,s_+)\delta x=1+Q_i\,\delta x,
$$
where $i\varphi$ and $\zeta_\pm$ are local random fields,
which
in terms of the random potential
$V(x)$
have the form
$\varphi(x)=-V(x)/k$,
$\zeta_\pm(x)=\pm i V(x)\, e^{\pm 2ikx}/2k$
(see~\cite{kol} for details).
Introducing the unitary matrices $U_i=(R)^i$ we can write
the transfer matrix for the loop chain in the following form:
\bea
T
&=&
R^N\prod_{i+1}^N U^*_i\,T_{i}\,U_i	\nonumber\\
&=&
R^N\prod_{i=1}^N
(1+U_i^*\,Q_i\,U_i\,\delta x)		\\
&=&
\exp(i\Theta L)\cdot
{\cal P}\exp
\int d\,x\;U^*(x)\,Q(x)\,U(x),		\nonumber
\eea
where $\cal P$ denotes the operator ordering
of the exponential
along the integration line.
Let us impose the resonant condition that
$\Theta$
be diagonal.
Choose $k$ so that
\beq
i\,\Theta =-ik\sigma_3,
\eeq
thus giving
\beq
U(x)=e^{-ikx\sigma_3}
\eeq
and
\beq\label{krr}
U^*(x)Q(x)U(x)={V(x)\over 2ik}
\left(
\matrix{1&-1\cr 1&-1\cr}
\right).
\eeq
The path-ordered exponential is easily computed since the matrix~\rf{krr}
is factorized:
\beq
T=
e^{-ikL\sigma_3}
\left(
\matrix{
1+{1\over 2ik}\int_{-L}^LV(x)\,d\,x
&-{1\over 2ik}\int_{-L}^LV(x)\,d\,x	\cr
{1\over 2ik}\int_{-L}^LV(x)\,d\,x
&1-{1\over 2ik}\int_{-L}^LV(x)\,d\,x	\cr
}
\right)
\eeq
It is worth noting that the non-exponential dependence of the matrix
$T$
on
$\int_{-L}^L\,V(t)\,dt$\/
follows from the nilpotency of the matrix~\rf{krr}.
The resistance of the loop chain is then
\bea\label{prima}
T_{12}T_{21}
&=&
{1\over 4k^2}\int dx\,dx'\, V(x)V(x').
\eea
This result is exact and does not depend on taking any average
(it refers of course to a very particular case).
For a random function $V(x)$ with space-homogeneous
statistics~\cite{lifshits_etal}
the right-hand side of~\rf{prima} is
a self-averaging quantity.
Therefore the bulk resistivity of the chain is equal to  $1/\ell$,
where
$\ell=4k^2/D$
is the localization length in the absence of loops.
We remember that in the absence of loops the resistance growths
exponentially with $L$.
The linear behaviour of the resistance (Ohm's law)
obtained in place of the exponential one shows that the localization
length diverges.

\begin{center}
\bf Acknowledgment
\end{center}

\noindent
We would like to thank Y.~Fedorov and I.B.~Khriplovich for helpful
advice and to V.~Benza for his interest in this work.

\end{document}